\documentclass[runningheads]{llncs}
\usepackage{graphicx}
\usepackage{comment}
\usepackage{amsmath,amssymb} 
\usepackage{color}
\usepackage{array}
\usepackage{siunitx}
\usepackage{algorithm}
\usepackage{algpseudocode}
\usepackage{algpascal}

\usepackage{xcolor}
\definecolor{blue}{rgb}{0, 0, 1.0}
\definecolor{green}{rgb}{0, .8, 0}
\definecolor{red}{rgb}{1.0, 0, 0}
\definecolor{black}{rgb}{0, 0, 0}
\definecolor{purple}{rgb}{0.5, 0, 0.5}
\newcommand{\hstz}[1]{\textcolor{blue}{#1}}

\newcommand*\samethanks[1][\value{footnote}]{\footnotemark[#1]}

\begin{document}
\pagestyle{headings}
\mainmatter
\def\ECCVSubNumber{4890}  

\title{Self-similarity Student for Partial Label Histopathology Image Segmentation} 

\titlerunning{Self-similarity Student for Partial Label Histopathology Image Segmentation}
%
\author{Hsien-Tzu Cheng\thanks{Both authors contributed equally to this work.}\inst{1} \and Chun-Fu Yeh\samethanks\inst{1} \and Po-Chen Kuo\inst{1,3} \and Andy Wei\inst{1} \and \\
Keng-Chi Liu\inst{1} \and Mong-Chi Ko\inst{1} \and Kuan-Hua Chao\inst{1} \and \\
Yu-Ching Peng\inst{2} \and Tyng-Luh Liu\inst{1,4}}

%
\authorrunning{H-T. Cheng et al.}
%

\institute{Taiwan AI Labs \and Taipei Veterans General Hospital 
\and National Taiwan University College of Medicine \and Institute of Information Science, Academia Sinica, Taiwan}

\maketitle

\begin{abstract}
Delineation of cancerous regions in gigapixel whole slide images (WSIs) is a crucial diagnostic procedure in digital pathology. This process is time-consuming because of the large search space in the gigapixel WSIs, causing chances of omission and misinterpretation at indistinct tumor lesions. To tackle this, the development of an automated cancerous region segmentation method is imperative.
We frame this issue as a modeling problem with partial label WSIs, where some cancerous regions may be misclassified as benign and vice versa, producing patches with noisy labels. To learn from these patches, we propose Self-similarity Student, combining teacher-student model paradigm with similarity learning. Specifically, for each patch, we first sample its similar and dissimilar patches according to spatial distance. A teacher-student model is then introduced, featuring the exponential moving average on both student model weights and teacher predictions ensemble. While our student model takes patches, teacher model takes all their corresponding similar and dissimilar patches for learning robust representation against noisy label patches. Following this similarity learning, our similarity ensemble merges similar patches' ensembled predictions as the pseudo-label of a given patch to counteract its noisy label. 
On the CAMELYON16 dataset, our method substantially outperforms state-of-the-art noise-aware learning methods by 5\% and the supervised-trained baseline by 10\% in various degrees of noise. Moreover, our method is superior to the baseline on our TVGH TURP dataset with 2\% improvement, demonstrating the generalizability to more clinical histopathology segmentation tasks.



\end{abstract}


\section{Introduction}
\begin{figure}[t!]
\begin{center}
\includegraphics[width=1.0\linewidth]{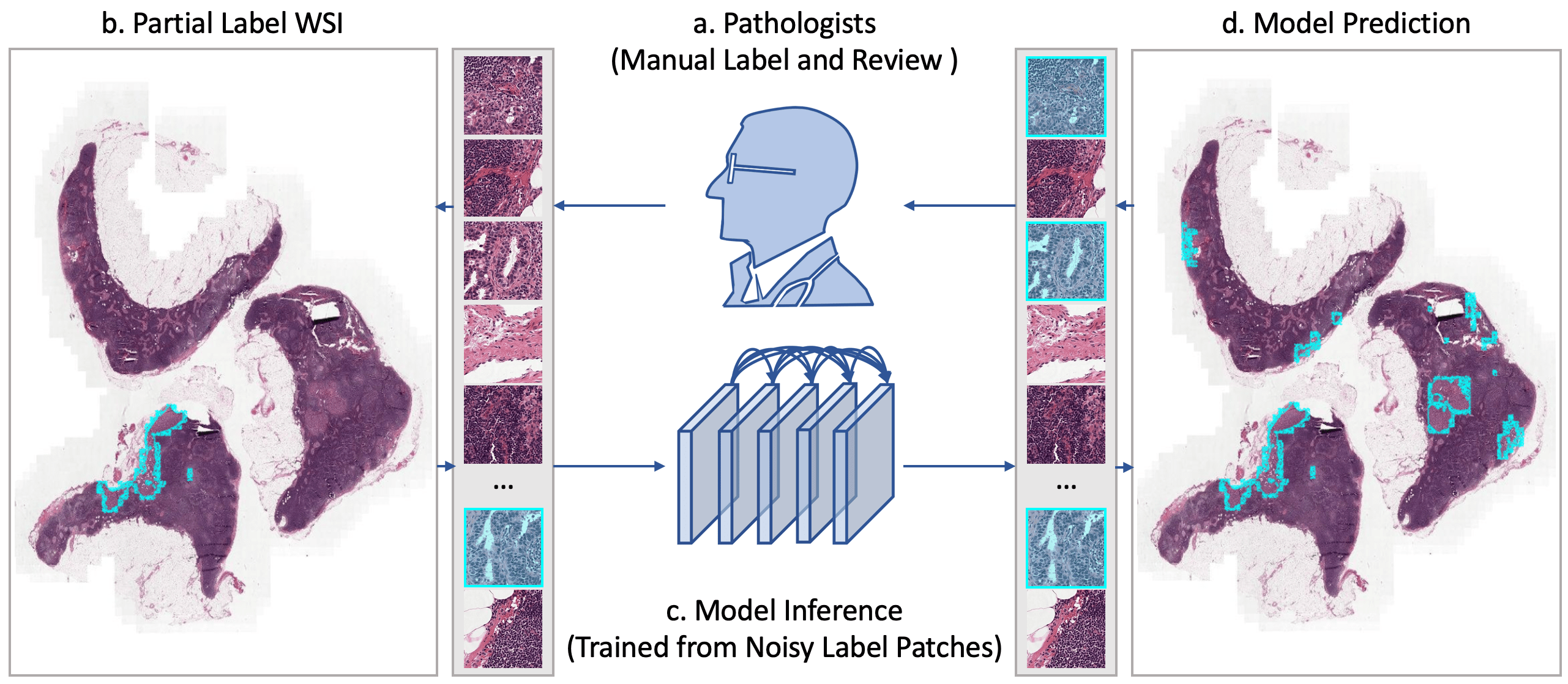}
\end{center}
\caption{
Overview of our application scenario. The cancerous regions and patches are masked by the cyan color. a. Pathologists manually annotate partial cancer regions in WSIs because of time constraints or misinterpretation. b. A partial label WSI with some lesions omission c. Our model inference with patches. d. The model output of patches combines to WSI for lesions prediction. This prediction can be used as a pseudo ground truth label for model training or for pathologists to review.}
\label{fig.overview}
\end{figure}

Digital pathology and deep learning (DL) techniques possess the potential to transform the clinical practice of pathology diagnosis. Conventionally, the gold standard for pathology diagnosis is the process of pathologists inspecting hematoxylin and eosin (H$\&$E) stained tissue specimens on glass slides using optical microscopes. This is time-consuming and error-prone. With the rising adoption of digital pathology, the digitized slide, namely whole slide image (WSI), mitigates the aforementioned issue. However, making diagnosis via manual inspection in gigapixel WSI is still labor intensive. The DL algorithms, particularly tailored to analyze WSI~\cite{komura2018machine}, would empower digital pathology and subsequently automate pathology diagnosis.


Identification of cancer regions in gigapixel WSI is considered the bottleneck of pathology diagnosis. Breaking this bottleneck, the CAMELYON16 challenge~\cite{bejnordi2017camelyon} serves as a milestone toward the automated segmentation of lymph node metastases in WSI. With detailed pixel-level annotation provided in this CAMELYON16 dataset, several fully-supervised DL algorithms have demonstrated the segmentation performance on par with pathologists~\cite{bejnordi2017camelyon,komura2018machine}. Still, amassing large scale WSI datasets for other diseases with the annotation comparable to CAMELYON16 requires a team of skilled pathologists and brings another bottleneck for the automation of digital pathology in clinical practice. To be scalable in clinics, DL algorithms, which leverage semi-supervised and weakly-supervised learning frameworks, may be effective.

To relieve the need of fine-grained annotations in WSI, semi-supervised and weakly-supervised learning frameworks deserve careful consideration. Recent work has successfully applied multiple instance learning (MIL) framework to detect cancer regions in WSI with weak labels (patient-level diagnosis)~\cite{campanella2019clinical,xu2019camel}. Nevertheless, research of the kind heavily depends on large datasets (more than 40,000 WSIs) to learn useful feature representations. This limits its applicability to the automation of digital pathology. Semi-supervised learning framework, on the other hand, demonstrates its potential to learn from datasets with part of them being completely or partially labeled \cite{peikari2018cluster,campanella2019clinical}. Such application scenario resembles the common clinical context, where pathologists miss small cancer regions and fail to identify the distinct boundary of tumor cells out of benign ones in WSI. Therefore, our paper aims to solve this issue with the technique originated from semi-supervised learning.

Inherent noises in partial label WSI may impede the learning ability of DL models. To alleviate the negative influence of noisy labels, teacher-student learning paradigm \cite{tarvainen2017meanteacher}, common in semi-supervised learning, tends to be helpful. Different from most of the semi-supervised works applying this paradigm to generate pseudo ground truths for unlabeled samples, those pseudo ground truths can also be used to eliminate or counteract noisy samples \cite{nguyen2019self}. Motivated by such paradigm, we further propose an approach to tackle the inherent noises originated from the modeling process with partial label WSI. Fig.~\ref{fig.overview} illustrates the application scenario of our proposed method and the inherent noises our method attempts to mitigate.

To accelerate the automation of digital pathology and deal with the subsequent modeling issue from partial label WSI, our proposed teacher-student model features the following strategies and contributions: 

\begin{enumerate}
    \item We propose Self-similarity Student, a teacher-student based model embedded with self-similarity learning and similarity predictions ensemble, to recognize cancer lesions from noisy label patches in partial label WSIs.
    \item Our self-similarity learning approach is motivated by the nature of tissue morphology, learning representation with a similarity loss that enforces nearby patches in a WSI to be closer in feature space.
    \item Our similarity ensemble approach generates pseudo label of a given patch in the partial label WSI. The similarity ensembled pseudo label is updated based on the consensus between predictions ensemble of the patch and its nearby patches, making the pseudo label more robust.
    \item The result on the CAMELYON16 dataset shows that our Self-similarity Student method achieves more than 10\% performance boost compared to the supervised-trained baseline and more than 5\% to the best previous art.
    \item The result of our method shows 2\% improvement over the baseline on our TVGH TURP cancer dataset, demonstrating the generalizability to more clinical histopathology segmentation tasks.
\end{enumerate}

\vspace{-2mm}
\section{Related Work} \label{sec.related_work}
\vspace{-2mm}
We discuss the relevant literature in two aspects, namely, recent research efforts on designing automatic analysis techniques for digital histopathology, and deep learning methods dealing with the semi-supervised scenario or noisy data, each of which is closely related to the problem setting of our method.
\vspace{-1mm}
\subsubsection{Digital histopathology} Concerning the extremely large sizes of WSIs (around $100k \times 50k$~pixels), designing automatic and effective machine learning techniques for histopathological image analysis is much needed in clinical practice \cite{komura2018machine}. Introduced by \cite{hou2016patch}, a CNN-based model has been proposed for patch-wise WSI classification, following a count-based aggregation for WSI-level classification.
Lee and Paeng \cite{lee2018robust} further adopt CNNs, comprising a patch-level detector and a slide-level classifier, for WSI metastasis detection and pN-stage classification of breast cancer. In \cite{takahama2019multi}, Takahama et al. propose to explore the global information from semantic segmentation to enhance the local classification performance on WSIs. To achieve this goal, they establish a DNN model that combines a patch-based classification module and a whole slide segmentation module. Campanella et al. \cite{campanella2019clinical} develop a clinical-grade computational pathology framework that utilizes multiple instance learning (MIL) based deep learning techniques to carry out a thorough study over a cancer dataset of $44,732$ WSIs from $15,187$ patients. The MIL setting allows the convenience of skipping annotating WSIs at the pixel level and also yields good classification performance. For automated segmentation of cancer regions in gigapixel WSIs, several works have addressed the challenging problem with deep learning in either fully supervised \cite{tokunaga2019adaptive}, or weakly supervised setting \cite{xu2019camel}. By assuming the correlation between image features of cancer subtypes and image magnifications, Tokunaga et al. \cite{tokunaga2019adaptive} propose the adaptive weighting multi-field-of-view CNN to carry out semantic segmentation for pathological WSIs. 
More recently, Xu et al. \cite{xu2019camel} introduce the CAMEL framework to address histopathology image segmentation in a weakly-supervised manner. Driven by MIL-based label enrichment, their method requires only image-level labels of training data, and progressively predicts instance-level labels and then pixel-level labels. For the post-process to combine patches, \cite{lin2018scannet,lin2019fastscannet} proposed to combine patches in multiple level of overlapping for smoother and less noisy WSI lesion segmentation. To clearly demonstrate the robustness of our method on dealing with noisy labeled data, we only conduct basic post-process, combining non-overlapped patches to WSIs, in all of our experiments.

\vspace{-1mm}
\subsubsection{Noisy label and semi-supervised learning.} The issue of noisy labeling in histopathological imaging is a major concern. In practice, the unreliable labeling is unavoidable in that manually annotating huge-size WSIs beyond image-level is inherently a daunting task. In addition, noisy labeling could also result from partially annotating WSIs as we aim to address in this work. Le et al. \cite{le2019pancreatic} develop a {\em noisy label classification} (NLC) to predict regions of pancreatic cancer in WSIs. Their method leverages a small set of clean samples to yield a weighting scheme for alleviating the effect of noisy training data. The {\em self-ensemble label filtering} (SELF) introduced in \cite{nguyen2019self} first uses the training dataset-wise running averages of the network predictions to filter noisy labels and then applies semi-supervised learning to achieve model training. SELF is shown to outperform other noise-aware techniques across different datasets and architectures. On semi-supervised learning, the temporal ensembling framework introduced in \cite{laine2016taumodel} is a pioneering effort on proposing self-ensembling mechanism that uses ensemble predictions to improve the quality of predictions on unknown labels. Their method is demonstrated to achieve significant improvements on standard benchmark datasets such as CIFAR-10 and CIFAR-100. Xie et al. \cite{xie2019noisystudent} develop a self-training method that a teacher model is learned from the labeled ImageNet images and is used to annotate pseudo labels from extra unlabeled data. Then the augmented training data of labeled and pseudo labeled images are used to learn the student model, where noise is injected to achieve generalization. The roles of teacher and student are then switched and the learning process is repeatedly carried out to obtain the final model. Our work differs from these previous arts in the following: (a) pseudo-labels from teacher-student model are used to counteract noisy labels instead of assigning them to unlabeled samples. (b) our pseudo-label of each patch is generated from the consensus of predictions ensemble of its similar patches.

\vspace{-1mm}
\section{Methodology}
\vspace{-1mm}
In this section, we elaborate our Self-similarity Student method for cancerous region segmentation in partial label WSIs. The preliminaries and notations are firstly shown in Sec.~\ref{sec.method.notation}. Next, we provide an overview of our proposed algorithm in Sec.~\ref{sec.method.overview}. After that, the method to construct similarity embedding is introduced in Sec.~\ref{sec.method.construct_sim}. At last, we describe the details of Self-similarity Student for noisy label learning in Sec.~\ref{sec.method.self_sim_noisy}

\vspace{-1mm}
\subsection{Preliminaries and Notations}\label{sec.method.notation}
\vspace{-1mm}
 
\begin{figure}[t!]
\begin{center}
\includegraphics[width=1.0\linewidth]{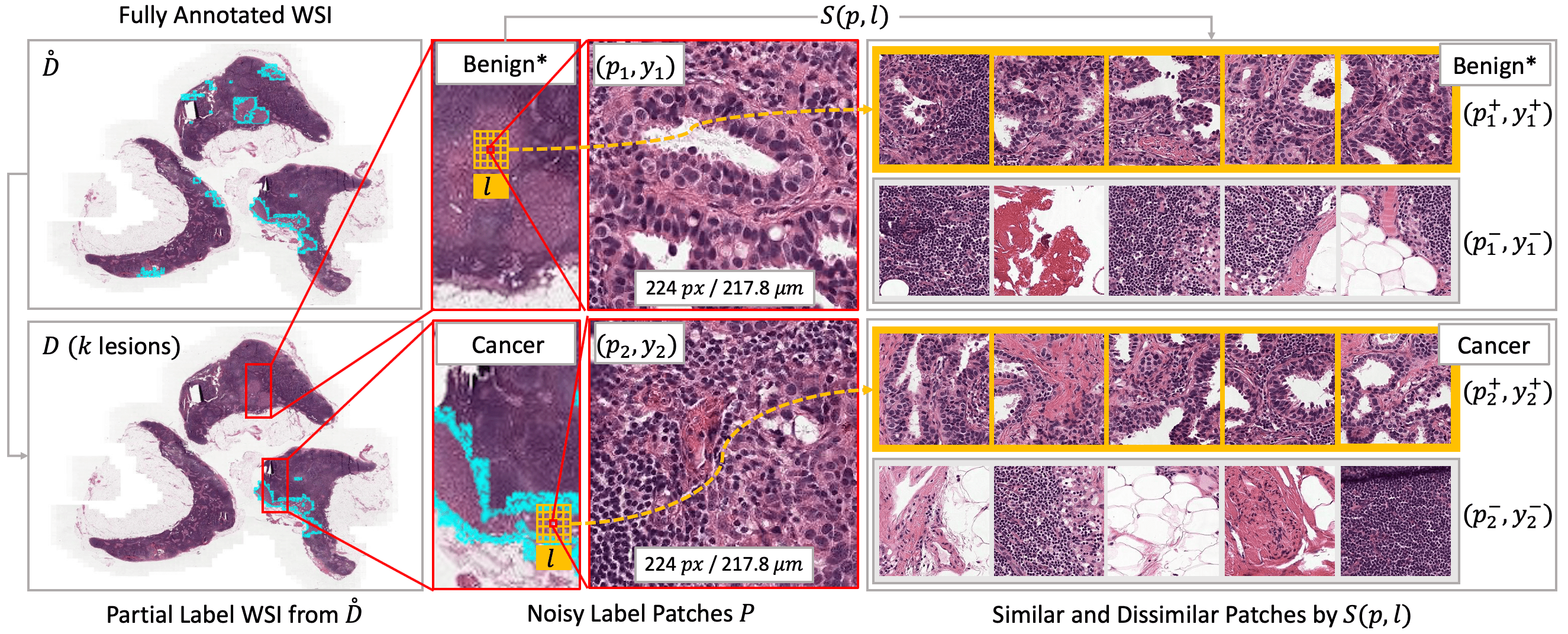}
\end{center}
\vspace{-1mm}
\caption{Similarity sampling. To simulate the scenario of training our model by only partial label WSIs, each $k$-lesion-remained $D$ is sampled from the $\mathring{D}$. From $D$ we sample noisy label patches $P$, illustrating 2 example patches $(p_1, y_1)$ and $(p_2, y_2)$. By the $\mathcal{S}(p,l)$ with distance threshold $l$ denoted in the orange grid, we can further sample the similar patches $(p^+, y^+)$ in orange box and dissimilar $(p^-, y^-)$ patches in gray box of each patch $p$.
The noisy label case occurs here because of the false benign (Benign*) label of $p_1$ which should be revised as Cancer by our model.
}
\label{fig.simsample}
\end{figure}

 Given a complete label WSI dataset $\{\mathring{D}_{train}, \mathring{D}_{val}, \mathring{D}_{test}\} = \mathring{D}$, we generate our partial label dataset $\{D_{train}, D_{val}\} = D$ for modeling and hold out clean test set $\mathring{D}_{test}$ for final evaluation. In partial label dataset $D$, k cancer lesions ($k_{top}$ or $k_{rand}$) per WSI are kept and the remaining ones are relabeled as non-cancerous regions. The $k_{top}$ and $k_{rand}$ stand for the top k largest and random k cancer lesions respectively. Moreover, patches $\{P_{train}, P_{val}\} = P $, which represent patch-label pairs $(p, y)$, are sampled from $D$. Since $D$ is partially annotated, label $y$ of each patch $p$ may be noisy. For each $(p, y)$ of a given WSI, we further sample its similar and dissimilar patches according to distance $l$, producing its similar $(p^+, y^+)$ and dissimilar $(p^-, y^-)$ bag of patches. The details of how we generate $D$ and $(p, y)$ are illustrated in Fig.~\ref{fig.simsample} and described in Sec.~\ref{sec.method.construct_sim}.

As to the teacher-student model, we consider $f$ to be a model with corresponding weights $\theta$ and augmentation $\eta$. Thus, we write the teacher model as $f_t(\theta_t, \eta_t)$ and student model as $f_s(\theta_s, \eta_s)$. Additionally, we denote the feature embedding of a given patch from teacher model and student model by $z_t$ and $z_s$. Similarly, the feature embeddings of similar and dissimilar pair ($p^+$ and $p^-$) for a given patch ($p$) from teacher model are expressed as $z_t^+$ and $z_t^-$ respectively.

\begin{figure}[h!]
\begin{center}
\includegraphics[width=1.0\linewidth]{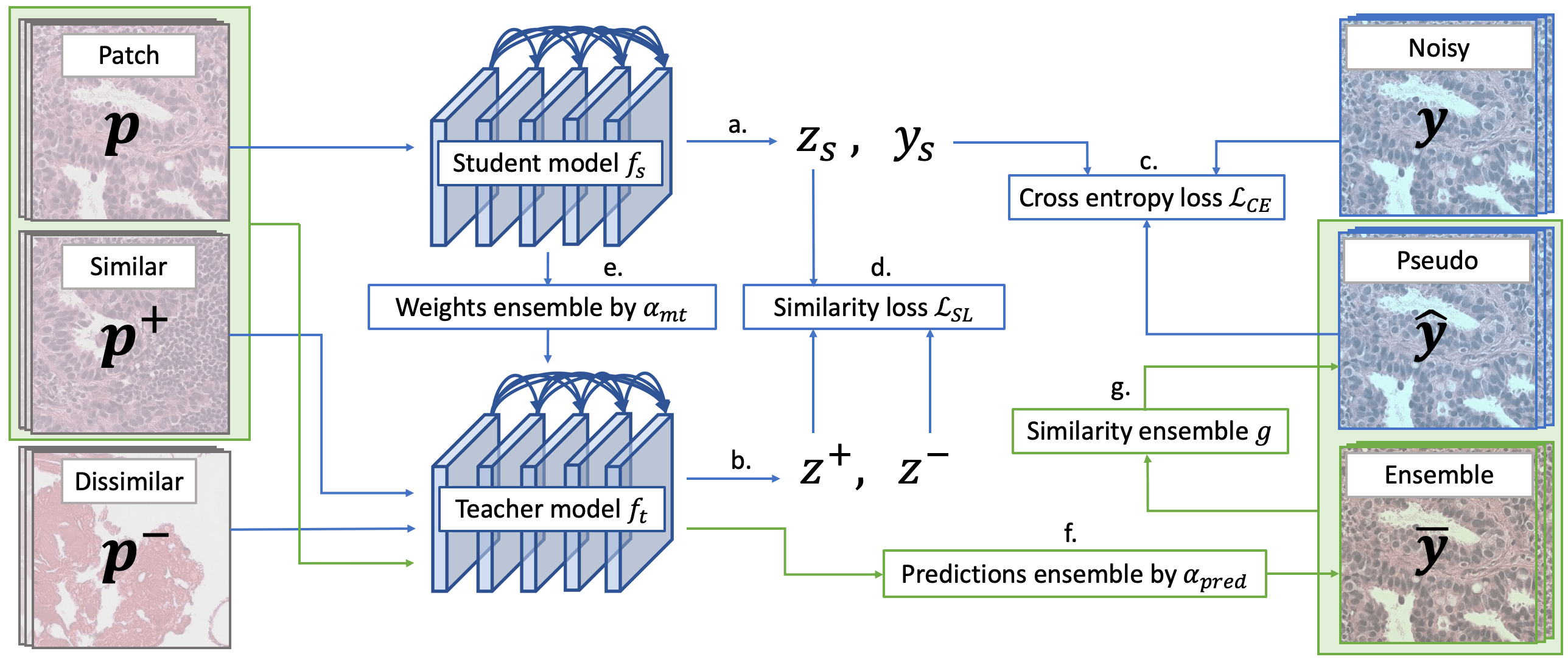}
\end{center}
\vspace{-1mm}
\caption{
Overview of our model training pipeline refers to Algorithm 1. For each epoch, we first train our student-teacher model (colored in blue) by means of $p$, $p^+$, $p^-$, with noisy label $y$ and pseudo label $\hat{y}$. $\mathcal{L}_{CE}$, $\mathcal{L}_{SL}$, and student weights EMA are calculated in every batch for the update. After $f_s$ and $f_t$ updated, we loop all patches again (colored in green), conducting EMA predictions ensemble then similarity ensemble to update our ensembled predictions $\bar{y}$ and pseudo label $\hat{y}$ which will be used in the next epoch.
}
\label{fig.method}
\end{figure}

\vspace{-1mm}
\subsection{Overview of Self-similarity Student}\label{sec.method.overview}
\vspace{-1mm}

Fig.~\ref{fig.method} illustrates an overview of our proposed approach. To deal with the noisy label patches $(p, y)$, our Self-similarity Student method builds upon two core concepts: teacher-student learning and similarity embedding. We build up our teacher and student models similar to the previous works \cite{tarvainen2017meanteacher,laine2016taumodel}. That is, the teacher model $f_t$ is the exponential moving average (EMA) of the student model $f_s$, which makes the intermediate representations more stable and facilitate student to learn robust representations against noisy samples. Besides the EMA of model weights for teacher model, we further leverage the ensemble of teacher model predictions for each patch to make the pseudo-label of each patch more consistent \cite{bachman2014pseudoensembles,nguyen2019self}. Different from  \cite{nguyen2019self} using predictions ensemble to filter noisy samples, the predictions ensemble $\bar{y}$ of each patch $p$ in our approach serves as the basis of its pseudo-labels $\hat{y}$ for complementary supervision, both counteracting noisy patches and making the most of original labels $y$ from $(p, y)$. 

In addition to the predictions ensemble, we introduce a similarity learning method to make the pseudo-labels of each patch more robust. This proposed method is motivated by the intrinsic property of tissue morphology in WSIs and the unsupervised visual representation learning work \cite{he2019moco}. Because nearby patches, which share similar morphological characteristics, tend to have the same labels, the generation of the pseudo-label $\hat{y}$ of a given patch $p$ could refer to its neighboring ones, namely $p^+$. Here, the pseudo-label $\hat{y}$ of a given patch $p$ is defined as a function $g(\bar{y}, \bar{y}^+)$. The $\bar{y}$ and $\bar{y}^+$ are the teacher predictions ensemble of a given patch $p$ and that of its similar pairs $p^+$ respectively. In this paper, we implement function $g$ as the average of $\bar{y}$ and $\bar{y}^+$. Following this intuition, we further apply similarity loss, which is a unique version of contrastive loss, to encourage the feature embeddings $z$ of nearby patches ($z$ and its $z^+$) to become closer and those of dissimilar ones ($z$ and its $z^-$) to be distinct in feature space. Details of the construction of similarity embedding by applying similarity loss are described in Sec.~\ref{sec.method.construct_sim} and in pseudo code provided in Algorithm 1.


\vspace{-1mm}
\alglanguage{pseudocode}
\begin{algorithm}[h!] 
\caption{Overview of our Self-similarity Student algorithm}
\begin{algorithmic}[0]
\Require $\{P_{train}, P_{val}\} = P$ \Comment{Noisy set of patches sampled from $D$}
\Require $\alpha_{mt}, \alpha_{pred} \in (0,1) \subset	\mathbb{R} $ \Comment{ EMA momentum}
\Require $l, ep_{max} \in \mathbb{N}$   \Comment{Distance threshold and max epoch}
\Require $\mathcal{O}$ = model weights gradient optimizer, e.g. Adam
\State Initialize $f_t(\theta_t, \eta_t)$  \Comment{Initialize teacher model}
\State Initialize $f_s(\theta_s, \eta_s)$  \Comment{Initialize student model}
\State $\hat{P} \gets P_{train}$ \Comment{Initialize all pseudo label $(p, \hat{y})$} 
\State $\bar{P} \gets P_{train}$ \Comment{Initialize all ensembled predictions $(p, \bar{y})$}
\ForAll {$(p, y) \in P_{train}$}
    \State $P_{train} \ni ((p^+, y^+), (p^-, y^-)) \gets \mathcal{S}(p, l)$  \Comment{Similarity sampling (Eq.~\ref{eq.sim_sample})}
\EndFor
\For{$ep \gets 0,\, ep_{max}$} \Comment{Main training loop}
    \ForAll {$(p, y) \in P_{train}$, $(p, \hat{y}) \in \hat{P}$}
        \State $ y_s, z_s \gets f_s(p)$
        \Comment{\textbf{a.} Student forward}
        \State $z_t^+ \gets f_t(p^+)$ \Comment{\textbf{b.} Teacher forward (Ignore $y_t^+$ and $y_t^-$)}
        \State  $z_t^- \gets f_t(p^-)$  
        \State $loss_{ce} \gets \mathcal{L}_{CE}(y, y_s) + \mathcal{L}_{CE}(\hat{y}, y_s)$ \Comment{\textbf{c.} Cross entropy loss (Eq.~\ref{eq.loss.crossentorpy})}
        \State $loss_{sl} \gets \mathcal{L}_{SL}(z_s, z_t^+, z_t^-)$ \Comment{\textbf{d.} Similarity embedding (Eq.~\ref{eq.loss.contrastive})}
        \State $\theta_s \gets  \mathcal{O} (\theta_s, loss_{ce}+loss_{sl})$ \Comment{Update student's weights} 
        \State $\theta_t \gets \alpha_{mt} \theta_t + (1-\alpha_{mt}) \theta_s$ \Comment{\textbf{e.} Update teacher's weights}
    \EndFor 
    \ForAll {$(p, \bar{y}) \in \bar{P}$} \Comment{\textbf{f.} Predictions ensemble}
        \State $\bar{y} \gets \alpha_{pred} \bar{y} + (1-\alpha_{pred}) f_t(p)$ \Comment{Update ensembled predictions per patch}
    \EndFor 
    \ForAll {$(p, \bar{y}) \in \bar{P}$, $(p, \hat{y}) \in \hat{P}$} \Comment{\textbf{g.} Similarity ensemble as pseudo label}
        \State $\hat{y} \gets g(\bar{y}, \bar{y}^+)$ \Comment{Consensus of nearby predictions}
    \EndFor 
\EndFor
\end{algorithmic}
\end{algorithm}
\vspace{-1mm}
\subsection{Construction of Similarity Embedding}\label{sec.method.construct_sim}
\vspace{-1mm}
Inspired by pathologists' empirical knowledge and the work \cite{gildenblat2019selfsim}, we formulate a similarity sampling strategy with respect to the distance $l$ between patches on a WSI, detailed in Fig.~\ref{fig.simsample}.
For each patch $(p,y)$, we sample multiple patches $p^+$ and $p^-$ by distance-based similarity sampling strategy $\mathcal{S}$ within a WSI:
\begin{equation} \label{eq.sim_sample}
  \mathcal{S}(p, l) = \begin{cases}
    p_i^+ \in p^+, i \in \{1, ..., N^+ \}, & \text{if $\|coord(p) - coord(p_i)\| \le l$},\\
    p_j^- \in p^-, j \in \{1, ..., N^- \}, & \text{otherwise}.
  \end{cases}
\end{equation}
\noindent where $coord$ denotes the coordinates of a patch on its original WSI and $N$ stands for the total number of patches in a WSI. Hence, there are $N^+ (< N)$ similar patches $p^+$ having Euclidean length to $p$ within distance $l$ and $N^- (= N - N^+$) dissimilar patches $p^-$ outside of that threshold.

To learn similarity embeddings in teacher-student models, a unique form of InfoNCE~\cite{oord2018contrastive} is considered in this paper. The formulation for similarity loss $\mathcal{L}_{SL}$ is as follows: 
\begin{equation}\label{eq.loss.contrastive}
    \mathcal{L}_{SL}(z_s, z_t^+, z_t^-) = -\log \frac{\sum_{i=1}^{N^+} \exp (z_s \cdot z_{t_i}^+ / \tau)}{\sum_{i=1}^{N^+} \exp (z_s \cdot z_{t_i}^+ / \tau) + \sum_{j=1}^{N^-} \exp (z_s \cdot z_{t_j}^-  / \tau)}
\end{equation}
where $\tau$~\cite{wu2018unsupervised} is a temperature hyper-parameter, $z_s$ is the student feature embedding for $p$, and $z_t^+$ and $z_t^-$ respectively are the teacher feature embeddings of its similar patches $p^+$ and dissimilar patches $p^-$ corresponding to $p$.

For time and memory efficiency to calculate $\mathcal{L}_{SL}$ in each iteration, we simplify equation~(\ref{eq.loss.contrastive}) by randomly sampling one patch from $p^+$ and that from $p^-$ for a given patch $p$ and further deriving $\mathcal{L}_{SL}$. This could be regarded as the log loss of a two-class softmax-based classifier, attempting to classify $z_s$ as $z_t$.
The dot products between the student feature embedding $z_s$ and the teacher feature embeddings, $z_t^+$ and $z_t^-$, can be viewed as the local similarity measurements.

\vspace{-1mm}
\subsection{Self-similarity Student for Noisy Label Learning}\label{sec.method.self_sim_noisy}
\vspace{-1mm}

In our proposed approach, the Self-similarity Student learns to both cluster local similar patches and classify each patch with the supervision from original labels $y$ and pseudo-labels $\hat{y}$. Unlike semi-supervised learning methods treating pseudo-labels from teacher-student paradigm as ground truths for unlabeled samples, the pseudo-labels $\hat{y}$ in our method are used to counteract noisy labels $y$ in $(p,y)$. With the similarity constraint mentioned in Sec.~\ref{sec.method.construct_sim}, a pseudo-label $\hat{y}$ of a given patch ${p}$, which is derived from the consensus between its predictions ensemble $\bar{y}$ and that $\bar{y}^+$ of its similar patches $p^+$, is guaranteed to be more stable and robust to noises.

Overall, we apply two categories of losses to encourage our Self-similarity Student to learn from noisy patch-label pairs $(p,y)$: $\mathcal{L}_{SL}$ and $\mathcal{L}_{CE}$. Specifically, $\mathcal{L}_{CE}$ includes the cross entropy loss between student predictions and original labels $y$, shown in equation~(\ref{eq.loss.crossentorpy}), and the cross entropy loss between student predictions and pseudo-labels $\hat{y}$.
\vspace{-2mm}
\begin{equation}\label{eq.loss.crossentorpy}
    \mathcal{L}_{CE}(y, f_s(p)) = -\log \sum_{k=0}^{N} y_k \log (f_s(p_k))\,.
\end{equation} 
\vspace{-2mm}
The overall lose function of our method can be written as
\begin{equation}\label{eq.loss.overall}
    \mathcal{L}_{overall} = \mathcal{L}_{CE}(y, f_s(p)) + \mathcal{L}_{CE}(\hat{y}, f_s(p)) + \mathcal{L}_{SL}(z_s, z_t^+, z_t^-)\,.
\end{equation} 
\vspace{-2mm}


\vspace{-1mm}
\section{Experimental Result}
\vspace{-1mm}
In this section, we first elaborate our experimental details about the implementation and dataset settings in Sec.\ref{sec.exp.imp_details} and Sec.~\ref{sec.exp.data_preprocess}. Next, the performance comparisons between our method and other variants are reported in Sec.~\ref{sec.exp.baseline_methods}, Sec.~\ref{sec.exp.various_label_ratio} are Sec.~\ref{sec.exp.ablation_study}. Lastly, Sec.~\ref{sec.exp.model_generality} shows the performance on the TVGH TURP dataset, suggesting the generalization potential of our method to other clinical WSI data. For performance evaluation, we use the dice similarity coefficient (DSC) as our patch-level metric and the free-response receiver operating characteristic (FROC) curve as our lesion-level metric. The FROC curve is defined as the plot of sensitivity versus the average number of false-positives per slide and the final FROC score is the average sensitivity at 6 predefined false positive rates, including $1/4$, $1/2$, $1$, $2$, $4$, $8$ false positives per slide~\cite{bejnordi2017camelyon}.





\vspace{-1mm}
\subsection{Implementation Details}\label{sec.exp.imp_details}
\vspace{-1mm}

We implement all baseline methods and our variants based on DenseNet121~\cite{huang2017densenet} baseline with its official ImageNet~\cite{deng2009imagenet} pretrained weight from PyTorch~\cite{paszke2019pytorch} model zoo. For optimizer $\mathcal{O}$ settings, we use Adam optimizer with learning rate 1e-4 and weight decay 4e-5, dividing learning rate by 2 per 50 epochs. The default random seed is set to 2020, dropout rate to 0.2, and batch size to 48 in all our experiments for fair comparison. 
Referred to common stain augmentations for WSIs~\cite{lee2018robust} and those used in robust mean-teacher based methods~\cite{cubuk2019randaugment,xie2019noisystudent}, we choose several augmentations to train all our models, as augmentations are proven crucial to the mean-teacher based model performance. For similarity sampling, we choose $l = 1 ~mm$ considering empirical cancer lesion diameter from pathologists' view. We also set $\alpha_{mt} = 0.999$, $\alpha_{pred} = 0.9$, $N^+ = 1$, $N^- = 1$, and $\tau =0.07$ in all our experiments. All our models are trained and tested for inference using one Nvidia RTX 2080 Ti GPU.

Here, we describe implementation details about the state-of-the-art baselines, as in Sec~\ref{sec.exp.baseline_methods}. We conduct similarity sampling on our method and ablation study only. For student network in Noisy Student~\cite{xie2019noisystudent}, we scale up the default dropout 2.5 times and all default augmentation functions 1.5 times. The teacher networks in both Mean Teacher~\cite{tarvainen2017meanteacher} are updated every batch by EMA of student's weights, while Predictions Ensemble~\cite{nguyen2019self} and Noisy Student are updated every epoch. We also change label filtering~\cite{nguyen2019self} to our relabel mechanism for balancing positive and negative samples. To the best of our knowledge, Mean Teacher, Predictions Ensemble, Noisy Student, and other relevant teacher-student methods are not designed for and experimented on noisy labeled histopathology WSIs yet.

\vspace{-1mm}
\subsection{Dataset}\label{sec.exp.data_preprocess}
\vspace{-1mm}
The CAMELYON16 dataset consists of 270 WSIs for training and 130 WSIs for testing. We randomly sample 243 WSIs as $\mathring{D}_{train}$ and 27 WSIs as $\mathring{D}_{val}$ from 270 WSIs. The 130 WSIs ($\mathring{D}_{test}$) are used for final evaluation. To simulate the clinical context of having unidentified cancer regions in WSI, we create partial label dataset $D_{train}$ and $D_{val}$ out of $\mathring{D}_{train}$ and $\mathring{D}_{val}$ by retaining $k_{top}$ and $k_{rand}$ cancer regions in each WSI. As shown in Table~\ref{tab.data_preprocess}, we choose the number of $k$ to be 1, 2, and 3. For example, $k_{top}=1$ means that only the largest caner lesion in each WSI is kept and the rest is regarded as benign tissue. It could be recognized that $k_{rand}$ tasks are more challenging than $k_{top}$ tasks due to the injected noises.

To further sample patches $p$ from WSIs in $D_{train}$ and $D_{val}$, we run OTSU thresholding~\cite{zhang2008image} and set a $50\%$ foreground-background ratio to extract foreground tissues. Following~\cite{le2019pancreatic}, cancerous patches $(p, y=1)$ are defined to have more than $50\%$ intersection of their areas with either $k_{top}$ or $k_{rand}$ cancerous regions, and benign patches $(p, y=0)$ are the ones fully from the area outside of those cancerous regions. The resulting patch-label pairs $(p,y) \in P$ with resolution $224 ~px \times 224 ~px$ are sampled from $10 \times$ magnification WSIs (0.972 $\mu m / px$), i.e. the receptive field of a patch covers $217.8 ~\mu m \times 217.8 ~\mu m$.




\begin{table}[h!] 
\caption{Number of patches in training set and statistics for various $k_{top}$ and $k_{rand}$ tasks. ``Complete''  denotes the task using the original $\mathring{D}$ with clean ground truth. We sample $P$ from $D$ for $k_{top}$ and $k_{rand}$ tasks. The ratio of noisiness declines as more lesions per WSI are correctly labeled. 
}
  \centering
    \small
    \begin{tabular}{|c|c|c|c|c|c|c|c|}
    \hline
    \multicolumn{1}{|c|}{} & \multicolumn{1}{c|}{} & \multicolumn{3}{c|}{$k_{top}$} & \multicolumn{3}{c|}{$k_{rand}$} \\
    \cline{3-8}
    Label & Complete & 1 & 2 & 3 & 1 & 2 & 3\\ 
    \hline
    Benign & 588286 & 597500 & 592454 & 590482 & 606342 & 604331 & 600122\\
    \hline
    Cancer & 21730 & 12516 & 17562 & 19534 & 2819 & 5188 & 8358 \\
    \hline
    \multicolumn{2}{|c|}{Ratio of Correct Cancer Patches} & 57.60\% & 80.82\% & 89.89\% & 12.97\% & 23.87\% & 38.46\% \\
    \hline
    \multicolumn{2}{|c|}{Ratio of Noisiness} & 42.40\% & 19.18\% & 10.11\% & 87.03\% & 76.13\% & 61.54\%\\
    \hline
    \multicolumn{2}{|c|}{Number of Cancer Lesions} & 91 & 150 & 193 & 91 & 150 & 193 \\
    \hline
    \multicolumn{2}{|c|}{Average Size of Lesions ($mm^2$)} & 6.7814 & 5.7527 & 4.9959 & 2.0587 & 1.8747 & 2.5277 \\
    \hline
    \end{tabular}%
  \label{tab.data_preprocess}%
\end{table}%

\vspace{-1mm}
\subsection{Comparison with Previous Arts} \label{sec.exp.baseline_methods}
\vspace{-1mm}
To benchmark our Self-similarity Student, we further implement several state-of-the-art methods, including Mean Teacher \cite{tarvainen2017meanteacher}, Noisy Student \cite{xie2019noisystudent}, and Predictions Ensemble \cite{nguyen2019self}. Demonstrated in Table.~\ref{tab.baseline_methods}, the results of our method outperform previous arts evaluated on $k_{top}=1$ and $k_{rand}=1$ tasks. Our method achieves 93.76 DSC \& 36.9 FROC on the $k_{top}=1$ task, and 85.56 DSC \& 31.88 FROC on the more challenging $k_{rand}=1$ task, which achieves more than 10\% performance boost compared to the supervised-trained baseline and more than 5\% performance boost compared to the best previous art.

\begin{figure}[h!]
\begin{center}
\includegraphics[width=1\linewidth]{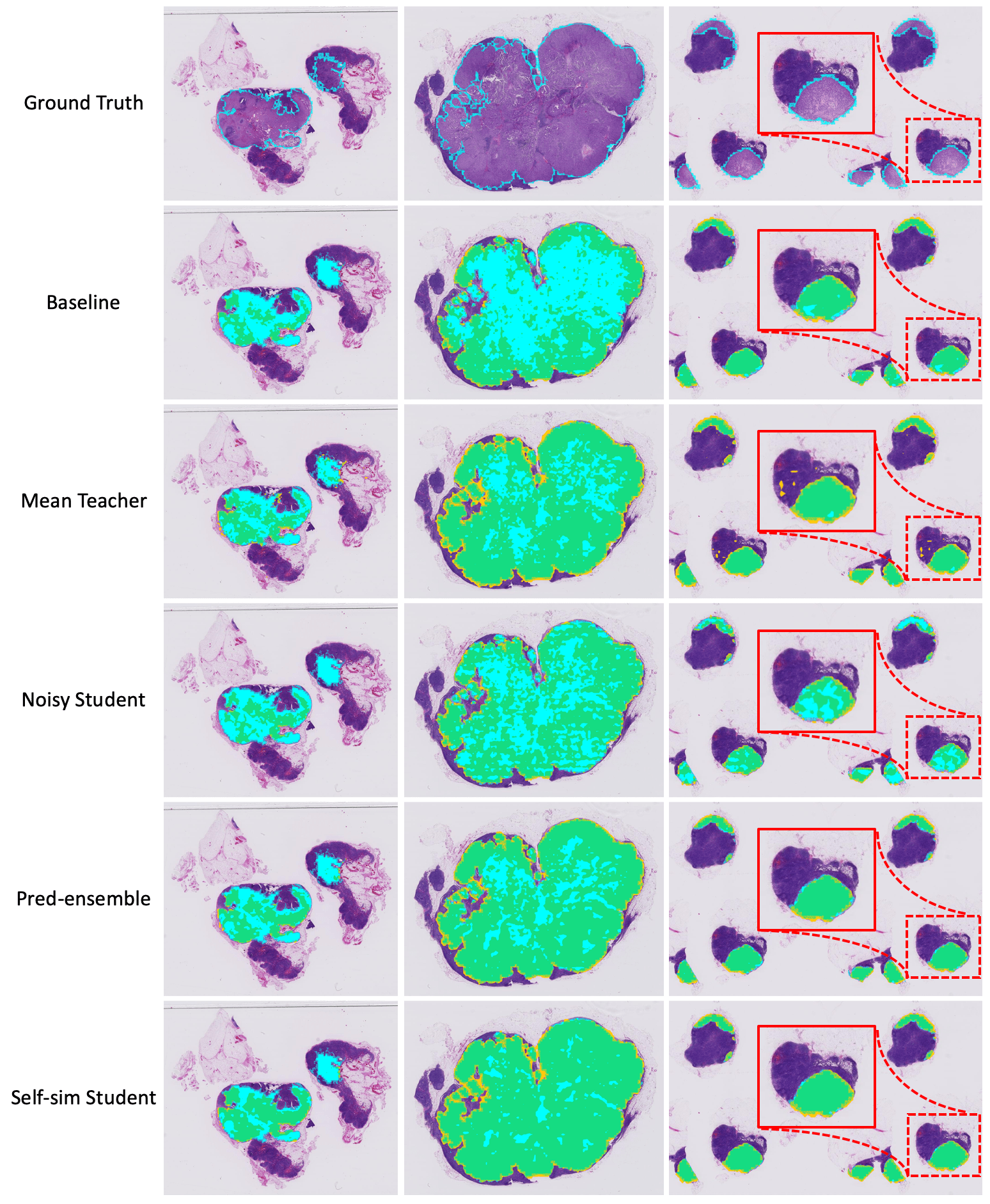}
\end{center}
\vspace{-1mm}
\caption{
Qualitative comparison with previous arts. Red boxes indicate zoomed-in region-of-interest. The regions in green color indicate true positives and yellow indicate false positives. The cyan color denotes ground truth (false negatives if no prediction overlapped). The top row demonstrates our method with lower false positives, especially compared to Mean Teacher. The lower two rows show that our method achieves higher sensitivity compared to all other baselines.}
\label{fig.qual_previous}
\end{figure}

Moreover, Fig.~\ref{fig.qual_previous} shows the $k_{top}=1$ qualitative comparison between our method and the baselines. Inferred from these results, our Self-similarity Student could both correctly identify more cancer regions and cancer cells (patches) with only a few false positives.
For the implementation details of baseline methods, see Supp. Sec. 2. For the illustration of effectiveness of our self-similarity embedding method, see Supp. Sec. 3. For more qualitative results, see Supp. Sec. 4.


\begin{table}[h!] 
\caption{Comparison with Previous Arts. All results are trained with single cancer lesion per WSI and evaluated on the clean testing set $\mathring{D}_{test}$. Our method outperforms other techniques in $k_{top}=1$ task (only one largest cancer region is annotated per WSI) and the more challenging $k_{rand}=1$ task (only one random cancer region is annotated per WSI)}
\vspace{-1mm}
  \centering
    \small
    \begin{tabular}{|c|c|c|c|c|}
    \hline
    \multicolumn{1}{|c|}{} & \multicolumn{2}{c|}{$k_{top}=1$} & \multicolumn{2}{c|}{$k_{rand}=1$} \\
    \cline{2-5}
    Method  & DSC & FROC & DSC & FROC \\ 
    \hline
    Baseline DenseNet121 \cite{huang2017densenet} & 83.08 & 29.99 & 63.08 & 28.09 \\
    \hline
    Mean Teacher \cite{tarvainen2017meanteacher} & 86.83 & 34.13 & 74.45 & 28.45 \\
    \hline
    Noisy Student \cite{xie2019noisystudent} & 84.90 & 34.21 & 77.46 & 30.06 \\
    \hline
    Prediction Ensemble \cite{nguyen2019self} & 88.60 & 33.41 & 75.59 & 30.20 \\
    \hline
    Self-sim Student \textbf{(Ours)}  & \textbf{93.76} & \textbf{36.90} & \textbf{85.56} & \textbf{31.88} \\
    \hline
    \end{tabular}%
  \label{tab.baseline_methods}%
\end{table}%

\begin{figure}[t!]
\begin{center}
\includegraphics[width=1\linewidth]{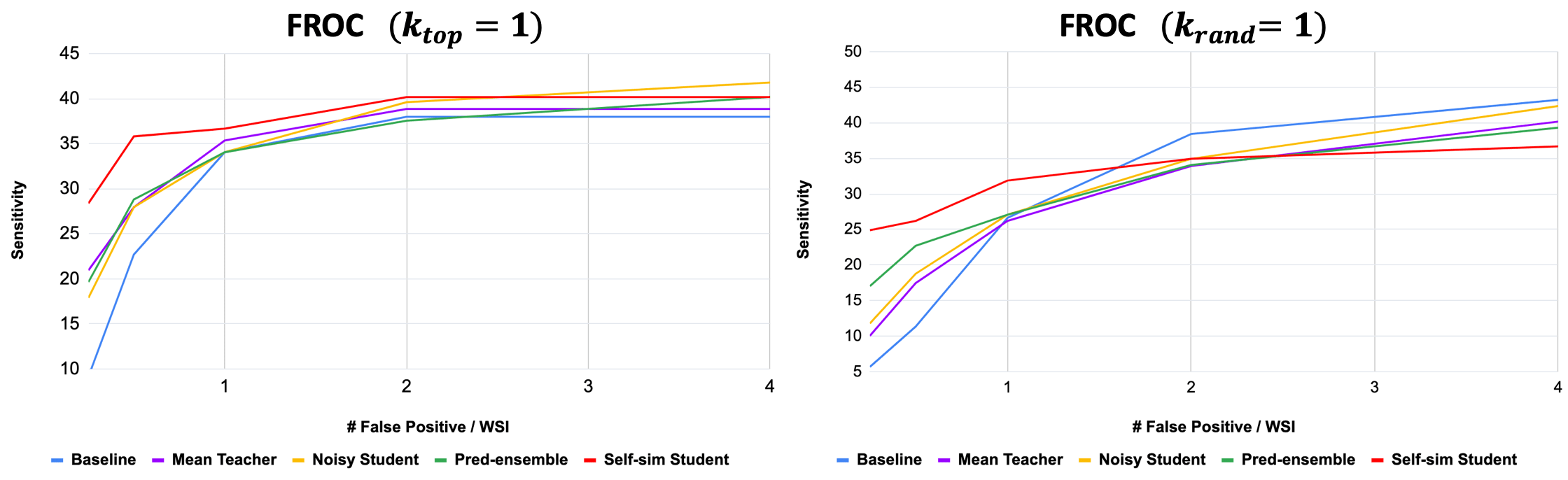}
\end{center}
\vspace{-1mm}
\caption{
FROC result. Our method, colored in red, yields the highest FROC score compared to baselines. Moreover, at the relatively low false positive rate, our method outperforms all the baselines with higher sensitivity.}
\label{fig.froc}
\end{figure}

\vspace{-1mm}
\subsection{Comparison with Various Label Ratio}\label{sec.exp.various_label_ratio}
\vspace{-1mm}
In this section, we compare the performances of the models trained from partially labeled training set described in Sec.~\ref{sec.exp.data_preprocess}. The results in Table~\ref{tab.baseline_methods} indicate that our method achieves comparable or better performance than DenseNet121 baseline in all experimental settings. These further suggest that our Self-similarity Student can still learn to discriminate most of the cancer regions from benign parts even in the situation, where there are $50\%$ cancer regions unidentified in training set.


\begin{table}[h!] 
\vspace{-1mm}
\caption{Performance of models trained in various noisy label ratios. Complete denotes the results trained from original ground truths. All results are from the clean testing set, demonstrating that our model is capable of learning from limited noisy ground truth while still having competent performance.} 
  \centering
    \small
    \begin{tabular}{|c|c|c|c|c|c|c|c|c|}
    \hline
    \multicolumn{1}{|c|}{} &  \multicolumn{1}{c|}{} & \multicolumn{1}{c|}{} & \multicolumn{3}{c|}{$k_{top}$} & \multicolumn{3}{c|}{$k_{rand}$} \\
    \cline{4-9}
    Method & Metrics & Complete & 1 & 2 & 3 & 1 & 2 & 3\\ 
    \hline

    Baseline & DSC & 92.68 & 83.08 & 87.40 & 90.41 &63.08 & 70.02 & 78.39 \\
    \cline{2-9}
    DenseNet121 \cite{huang2017densenet} & FROC & 41.12 & 29.99 & 38.37 & \textbf{40.18} & 28.09 & 35.90 & 33.22\\
    \hline

    Self-sim Student & DSC & 90.49 & \textbf{93.76} & \textbf{93.98} & \textbf{91.29} & \textbf{85.56} & \textbf{87.57} & \textbf{90.20} \\
    \cline{2-9}
    \textbf{(Ours)} & FROC & 39.52 & \textbf{36.90} & \textbf{38.94} & 39.16 & \textbf{31.88} & \textbf{36.54} & \textbf{35.08} \\
    \hline
    \end{tabular}%
  \label{tab.baseline_methods}%
\end{table}%
\vspace{-1mm}
\subsection{Ablation Study}\label{sec.exp.ablation_study}
\vspace{-1mm}
To demonstrate the contributions by each part of our Self-similarity Student method, we conduct a set of experiments in $k_{rand}=1$ task. Baseline indicates the DenseNet121 network trained with $\mathcal{L}_{CE}(y,\acute{y})$ whereas Sim-embedding shows the performance of Baseline plus $\mathcal{L}_{SL}$. Furthermore, Pred-ensemble uses the teacher predictions ensemble of each patch as its pseudo label while Sim-ensemble derives pseudo-labels by averaging the predictions ensemble from patches and their corresponding similar pairs. 

Shown in Table.~\ref{tab.ablation_Study}, each component in Self-similarity Student contributes to its overall performance. Most importantly, this justifies the effectiveness of our method, which makes pseudo-labels more robust against noisy labels by using predictions ensemble from similar pairs.


\begin{table}[h!] 
\caption{Ablation Study. Abbreviations: Sim-ensemble stands for similarity ensemble; Sim-embedding stands for similarity embedding using loss learning; Pred-ensemble stands for predictions ensemble.}
  \centering
    \small
    \begin{tabular}{|cccc|c|c|}
    \hline
    \multicolumn{4}{|c|}{Ablation Study} & \multicolumn{1}{c|}{DSC} & \multicolumn{1}{c|}{FROC} \\
    \cline{1-4}
    \multicolumn{1}{|c|}{Baseline} & \multicolumn{1}{c|}{Sim-embedding} & \multicolumn{1}{c|}{Pred-ensemble} & \multicolumn{1}{c|}{Sim-ensemble} & & \\ 
    \hline
    \checkmark & & & & 63.08 & 28.09 \\
    \hline
    \checkmark & \checkmark & & & 68.42 & 28.24 \\
    \hline
    \checkmark & & \checkmark &  & 75.59 & 30.20 \\
    \hline
    \checkmark & \checkmark & \checkmark & \checkmark & \textbf{85.56} & \textbf{31.88} \\
    \hline
    \end{tabular}%
  \label{tab.ablation_Study}%
\end{table}%


\begin{figure}[h!]
\begin{center}
\includegraphics[width=1\linewidth]{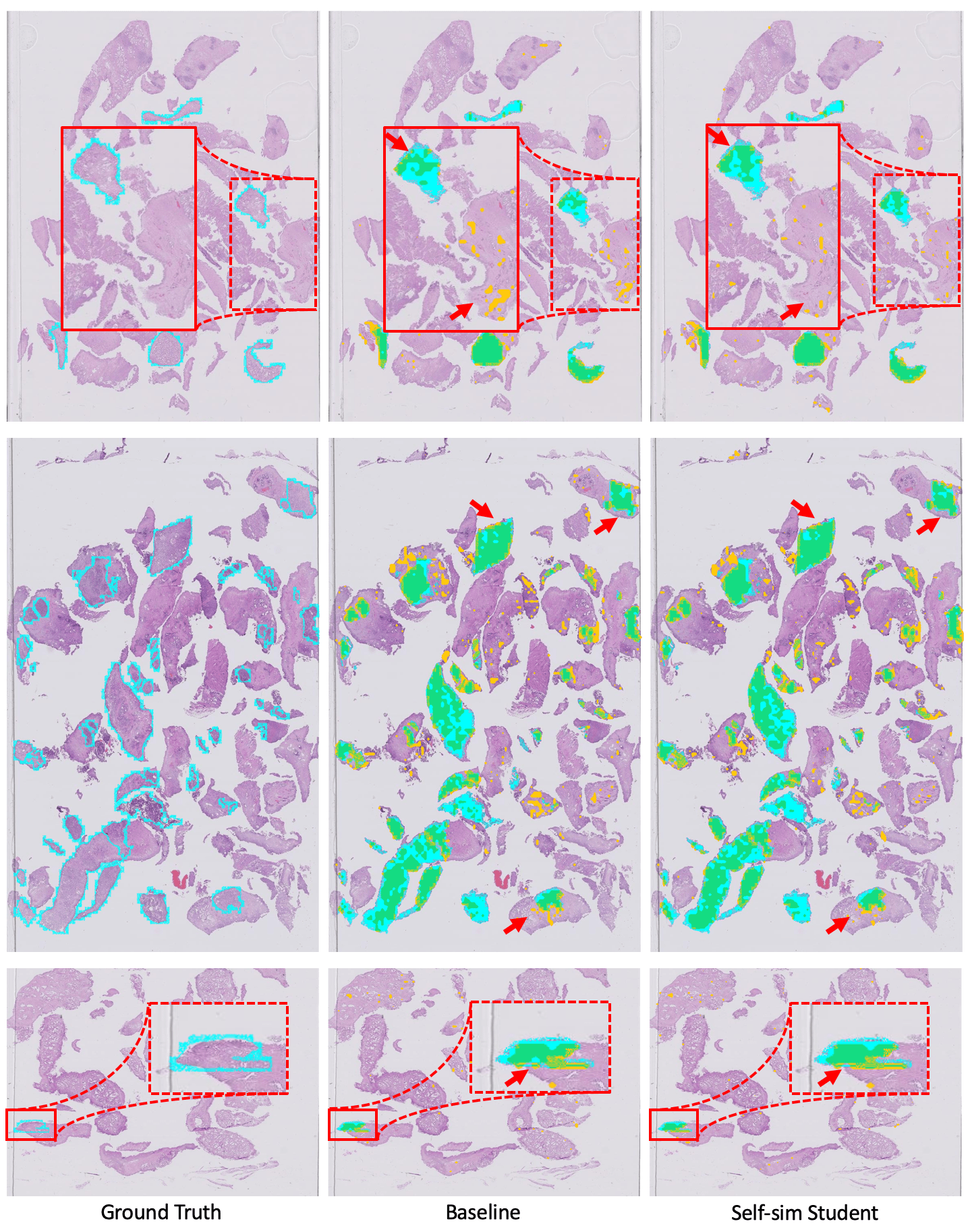}
\end{center}
\vspace{-1mm}
\caption{
Qualitative result on the testing set of TVGH TURP dataset. The color code is the same as Fig.~\ref{fig.qual_previous}. 
Our Self-similarity Student is able to predict cancer regions more precisely than the baseline with better grouped patterns.}
\label{fig.tvgh_turp}
\end{figure}

\vspace{-1mm}
\subsection{Generalizability of our method}\label{sec.exp.model_generality}
\vspace{-1mm}
To evaluate the potential generalizability of our method, we conduct experiments on the transurethral resection of prostate (TURP) data from the Department of Pathology, Taipei Veterans General Hospital (TVGH). The TVGH TURP dataset consists of 71 WSIs with annotated cancerous lesions, defined as regions with Gleason score greater than $3+3$. The training set consists of 58 WSIs within 13 WSIs for validation. The actual size of each WSI is about $45.7 mm \times 24.7 mm$ with 0.25 $\mu m/px$ pixel spacing. We follow the same patch extraction strategy in CAMELYON16 dataset, producing totally 459273 patches from $4 \times$ zoomed TURP WSIs with receptive field $224 \times 224 \mu m$. Our method achieves 77.24 DSC on the 13 WSIs testing set, which is better than the supervise-trained baseline with 75.36 DSC. The qualitative result is shown in Fig.~\ref{fig.tvgh_turp} and Supp. Sec. 4.

\vspace{-3mm}
\section{Conclusion}
\vspace{-3mm}
Computer vision technique is the key to accelerate the automation of digital pathology in clinical practice, particularly the identification of cancer regions in WSI. In this research, we propose a teacher-student framework, Self-similarity Student, to address partial label WSI, which relieves the burden on pathologists. The result shows that our method outperforms previous arts at least $5\%$ in terms of DSC, suggesting that Self-similarity Student possesses more robust representations against noisy labels. Following these meticulous experiments, the advantage of similarity ensemble for modeling with partial label WSI is verified. More importantly, our approach is capable of generalizing to TURP dataset, which identifies cancer regions out of benign ones with fewer false positives. To sum, Self-similarity Student can be a potent method to tackle the problems originated from partial label WSI dataset. In future work, we aim to pursue the potential of Self-similarity Student for semi-supervised learning in small-sized WSI datasets, truly contributing to the automation of cancer region delineation.

\subsubsection{Acknowledgment.}
We thank Yi-Chin Tu, the chairman of Taiwan AI Labs, for the generous support of this project. We also thank the intensive assistance made by the Department of Pathology and Laboratory Medicine, Taipei Veterans General Hospital. Lastly, we appreciate Tsun-Hsiao Wang at National Yang-Ming University for his contribution on delineating cancerous regions in WSIs of TVGH TURP dataset.


\newpage
\bibliographystyle{splncs04}


\newpage
\title{Supplementary Material: \\
    Self-similarity Student for Partial Label Histopathology Image Segmentation}
\author{Hsien-Tzu Cheng\thanks{Both authors contributed equally to this work.}\inst{1} \and Chun-Fu Yeh\samethanks\inst{1} \and Po-Chen Kuo\inst{1,3} \and Andy Wei\inst{1} \and \\
Keng-Chi Liu\inst{1} \and Mong-Chi Ko\inst{1} \and Kuan-Hua Chao\inst{1} \and \\
Yu-Ching Peng\inst{2} \and Tyng-Luh Liu\inst{1,4}}

\authorrunning{H-T. Cheng et al.}

\institute{Taiwan AI Labs \and Taipei Veterans General Hospital 
\and National Taiwan University College of Medicine \and Institute of Information Science, Academia Sinica, Taiwan}

\maketitle
\section{Augmentations for Model Training} \label{sec.augmentation}

We apply augmentations $\eta_t$ and $\eta_s$ for training teacher-studnet models \cite{cubuk2019randaugment,lee2018robust,xie2019noisystudent} on partial label WSIs. Several augmentations are chosen as shown in Table.~\ref{tab.augmentation}. For network dropout, we set all teacher models' dropout to 0.0 whereas student models' following Table.~\ref{tab.baselines}. We use Noisy augmentation hyperparameters to train the student model in Noisy Student, and Normal augmentation hyperparameters to train the other teacher-student models.


\begin{table}[h!] 
\caption{Augmentations.}
  \centering
    \small
    \begin{tabular}{|c|c|c|c|}
    \hline
    \multicolumn{2}{|c|}{} & \multicolumn{2}{c|}{Hyperparameters} \\
    \cline{3-4}
    \multicolumn{2}{|c|}{Augmentations} & Normal & Noisy \\ 
    \hline
      & Contrast (delta) & (0.75, 1.25) & (0.5, 1.875) \\ 
    \cline{2-4}
    & Brightness (delta) & (-0.2, 0.2) & (-0.3, 0.3) \\
    \cline{2-4}
    & Hue (delta) & (-0.05, 0.05) & (-0.075, 0.075)\\
    \cline{2-4}
    Stain/Color & Saturation (delta) & (0.8, 1.2) & (0.533, 1.8)\\
    \hline
     & Flip (probability) & \multicolumn{2}{c|}{0.5}\\
    \cline{2-4}
     & Resize (scale) & (0.9, 1.1) & (0.6, 1.35) \\
    \cline{2-4}
     & Crop (resolution) & \multicolumn{2}{c|}{(224, 224)}\\
    \cline{2-4}
     & Rotation (degree) & \multicolumn{2}{c|}{(-180, 180)}\\
    \cline{2-4}
    Deformation & Translation (delta) & (-0.05, 0.05) & (-0.075, 0.075)\\
    \hline
    Network & Dropout (ratio)& 0.2 & 0.5\\
    \hline
    \end{tabular}%
  \label{tab.augmentation}%
\end{table}%

\alglanguage{pseudocode}
\begin{algorithm}[h!] 
\caption{Teacher-student Model}
\begin{algorithmic}[0]
\Require $\{P_{train}, P_{val}\} = P$ \Comment{Noisy set of patches sampled from $D$}
\Require $\alpha^b_{mt}, \alpha^e_{mt}, \alpha_{pred}, \lambda \in (0,1) \subset	\mathbb{R} $ \Comment{ EMA momentum and $\mathcal{L}_{cs}$ weight}
\Require $ep_{max} \in \mathbb{N}$   \Comment{Distance threshold and max epoch}
\Require $\mathcal{O}$ = model weights gradient optimizer, e.g. Adam
\State Initialize $f_t(\theta_t, \eta_t)$  \Comment{Initialize teacher model}
\State Initialize $f_s(\theta_s, \eta_s)$  \Comment{Initialize student model}
\State $\hat{P} \gets P_{train}$ \Comment{Initialize all pseudo label $(p, \hat{y})$} 

\For{$ep \gets 0,\, ep_{max}$} \Comment{Main training loop}
    \ForAll {$(p, \hat{y}) \in \hat{P}$}
        \State $ y_s, z_s \gets f_s(p)$ \Comment{Student forward}
        \State $ y_t, z_t \gets f_t(p)$ \Comment{Teacher forward}
        \State $loss_{ce} \gets \mathcal{L}_{CE}(\hat{y}, y_s)$ \Comment{Cross entropy loss}
        \State $loss_{cs} \gets \mathcal{L}_{CS}(z_s, z_t)$ \Comment{Consistency loss (Eq.~\ref{eq.loss.constist})}
        \State $\theta_s \gets  \mathcal{O} (\theta_s, loss_{ce}+\lambda loss_{cs})$ \Comment{Update student's weights} 
        \State $\theta_t \gets \alpha^b_{mt} \theta_t + (1-\alpha^b_{mt}) \theta_s$ \Comment{Update teacher's weights per batch}
    \EndFor 
    \State $\theta_t \gets \alpha^e_{mt} \theta_t + (1-\alpha^e_{mt}) \theta_s$ \Comment{Update teacher's weights per epoch}
    \ForAll {$(p, \hat{y}) \in \hat{P}$} \Comment{Predictions ensemble}
        \State $\hat{y} \gets \alpha_{pred} \hat{y} + (1-\alpha_{pred}) y_t$ \Comment{Update pseudo label}
    \EndFor 
\EndFor
\end{algorithmic}
\end{algorithm}
\begin{table}[t!] 
\caption{Hyper parameters of baseline previous arts.}
  \centering
    \small
    \begin{tabular}{|c|p{1cm}|p{1cm}|p{1cm}|p{1cm}|}
    \hline
     & $\alpha^b_{mt}$ & $\alpha^e_{mt}$ & $\alpha_{pred}$ & $\lambda$ \\
    \hline
    Mean Teacher & 0.999 & 1 & 0 & 1 \\
    \hline
    Noisy Student & 1 & 0 & 0 & 0 \\
    \hline
    Pred-ensemble & 0.999 & 1 & 0.9 & 0 \\
    \hline
    \end{tabular}%
  \label{tab.baselines}%
\end{table}%

\section{Implementation Details of Baseline Methods}\label{sec.baselines}



In this section, we describe more about the implementation of our baseline previous art methods. We implement Mean Teacher~\cite{tarvainen2017meanteacher}, Noisy Student \cite{xie2019noisystudent}, and Pred-ensemble \cite{nguyen2019self} for comparison with our Self-similarity Student method. The teacher-student model training pipeline is illustrated in Algorithm 1. In each epoch, both teacher model and student model inference the noisy label patch $p$. The supervised loss between the (pseudo) label $\hat{y}$ and student model prediction $y_s$ is the calculated by cross entropy loss ${L}_{CE}$. Following \cite{tarvainen2017meanteacher}, the consistency loss ${L}_{CS}$ (Eq.~\ref{eq.loss.constist}) is used to constrain the feature map outputs of teacher model $f_t$ and student model $f_s$ to have similar predictions:
\begin{equation}\label{eq.loss.constist}
\mathcal{L}_{CS}(z_t, z_s) = \|z_t-z_s\|^{2}
\end{equation}
where $z_t$ and $z_s$ are feature maps from fully-connected layers of teacher-student models.
After the optimizer $\mathcal{O}$ backpropagated the loss, model weights ensemble and predictions ensemble are applied according to different settings respectively. Table.~\ref{tab.baselines} shows all the hyperparameters settings of different previous art baselines. For per-batch EMA momentum $\alpha^b_{mt}$ and per-epoch EMA momentum $\alpha^e_{mt}$, 1 denotes no weights update and 0 denotes entirely weights replacement of teacher model from student model. For predictions ensemble momentum $\alpha_{pred}$, 1 denotes no label update and 0 denotes the entirely pseudo label update by the teacher model prediction. As indicated in Table.~\ref{tab.baselines}, following their original settings, we update Mean Teacher and Pred-ensemble every batch while update Noisy Student every epoch. We set $\alpha_{pred}=0.9$ for Pred-ensemble and only apply $\mathcal{L}_{CS}$ on Mean Teacher. Note that we use pseudo label mechanism, instead of eliminating patches by noisy label filtering, to stabilize the training process without overly filtering.

\begin{figure}[t!]
\begin{center}
\includegraphics[width=1\linewidth]{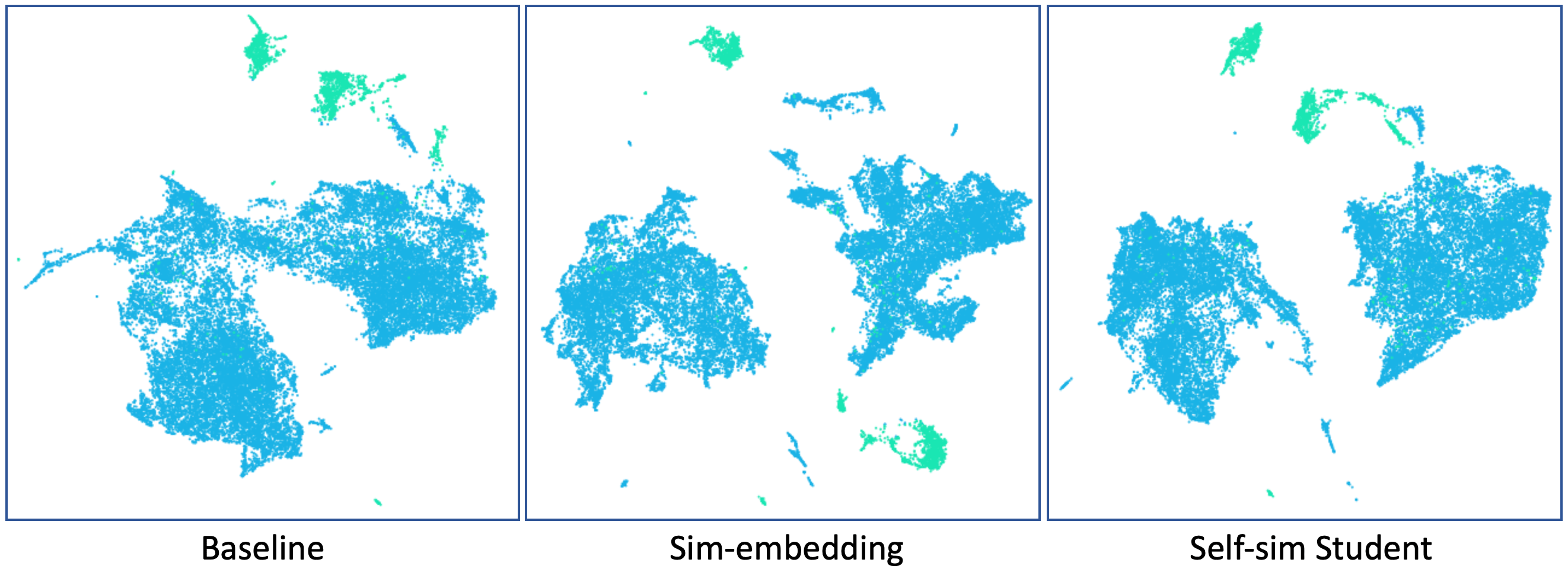}
\end{center}
\caption{
Qualitative result of the distribution of feature embeddings with UMAP. The cyan colored points denote benign patches and light-green colored points denote cancerous patches. Our method is able to learn a more compact feature representation for both benign and cancerous patches.}
\label{fig.sim_enb}
\end{figure}

\vspace{-1mm}
\section{Feature Embedding of Self-similarity Student} \label{sec.exp.sim_ens}
\vspace{-1mm}

To illustrate the effectiveness of similarity learning, we use UMAP~\cite{2018arXivUMAP} dimension reduction algorithm to visualize the feature embeddings derived from our method and the baselines. Specifically, we sample 30000 patches from our testing set $\mathring{P}$ and apply UMAP on the feature embeddings (n dimension=1024). As shown in Fig.~\ref{fig.sim_enb}, with similarity learning, Self-similarity Student can learn a more compact distribution of feature embeddings, which support its advantage in identifying cancerous patches over the baseline.


\section{Additional Qualitative Result}\label{sec.exp.imp_details}

\begin{figure} 
\begin{center}
\includegraphics[width=1\linewidth]{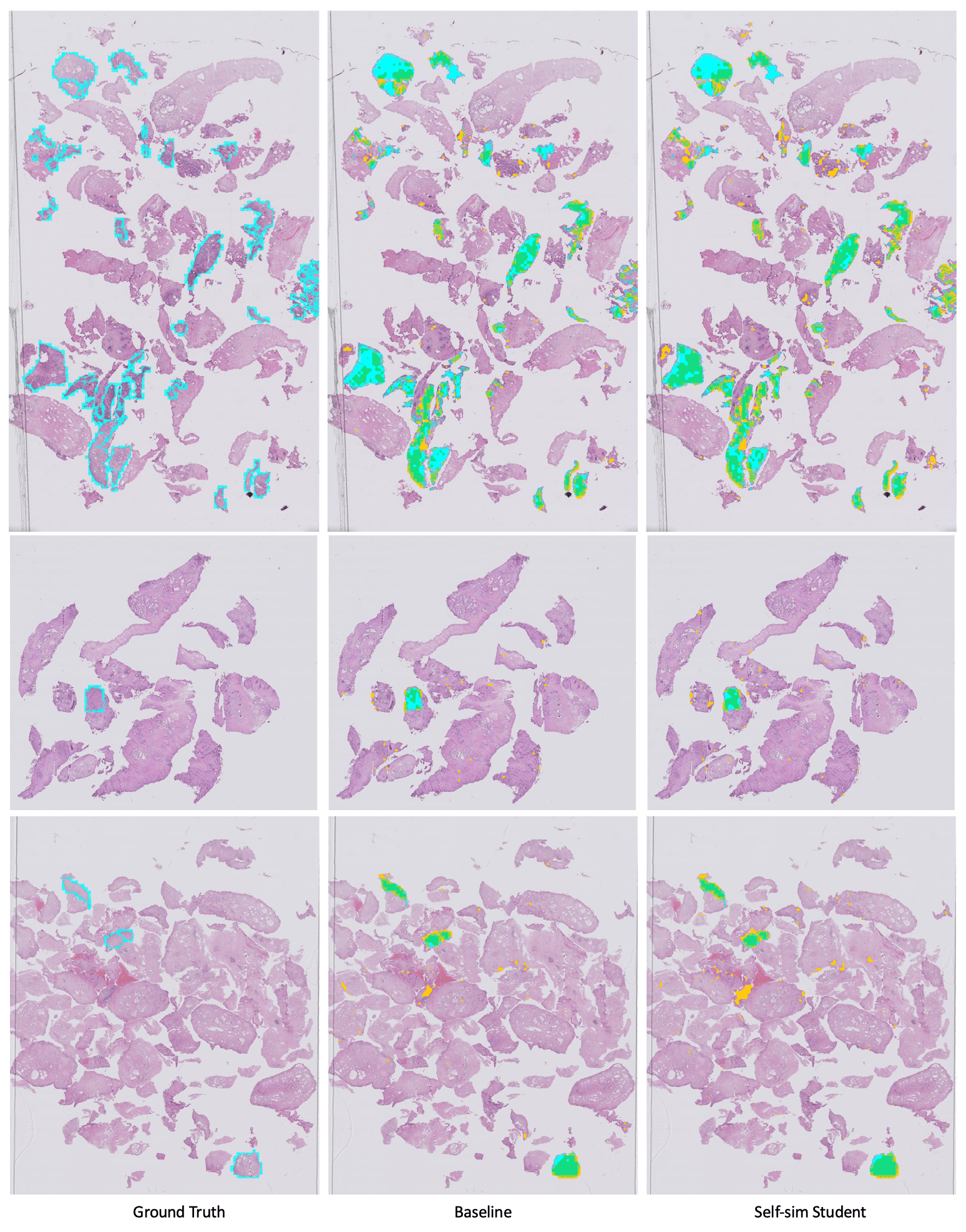}
\end{center}
\caption{
Additional qualitative result on TVGH TURP dataset. The regions in green color indicate true positives and yellow indicate false positives. The cyan color denotes ground truth (false negatives if no prediction overlapped). Our Self-similarity Student is able to predict cancer regions more precisely than the baseline with better grouped patterns.}
\label{fig.add_turp}
\end{figure}

\begin{figure} 
\begin{center}
\includegraphics[width=1\linewidth]{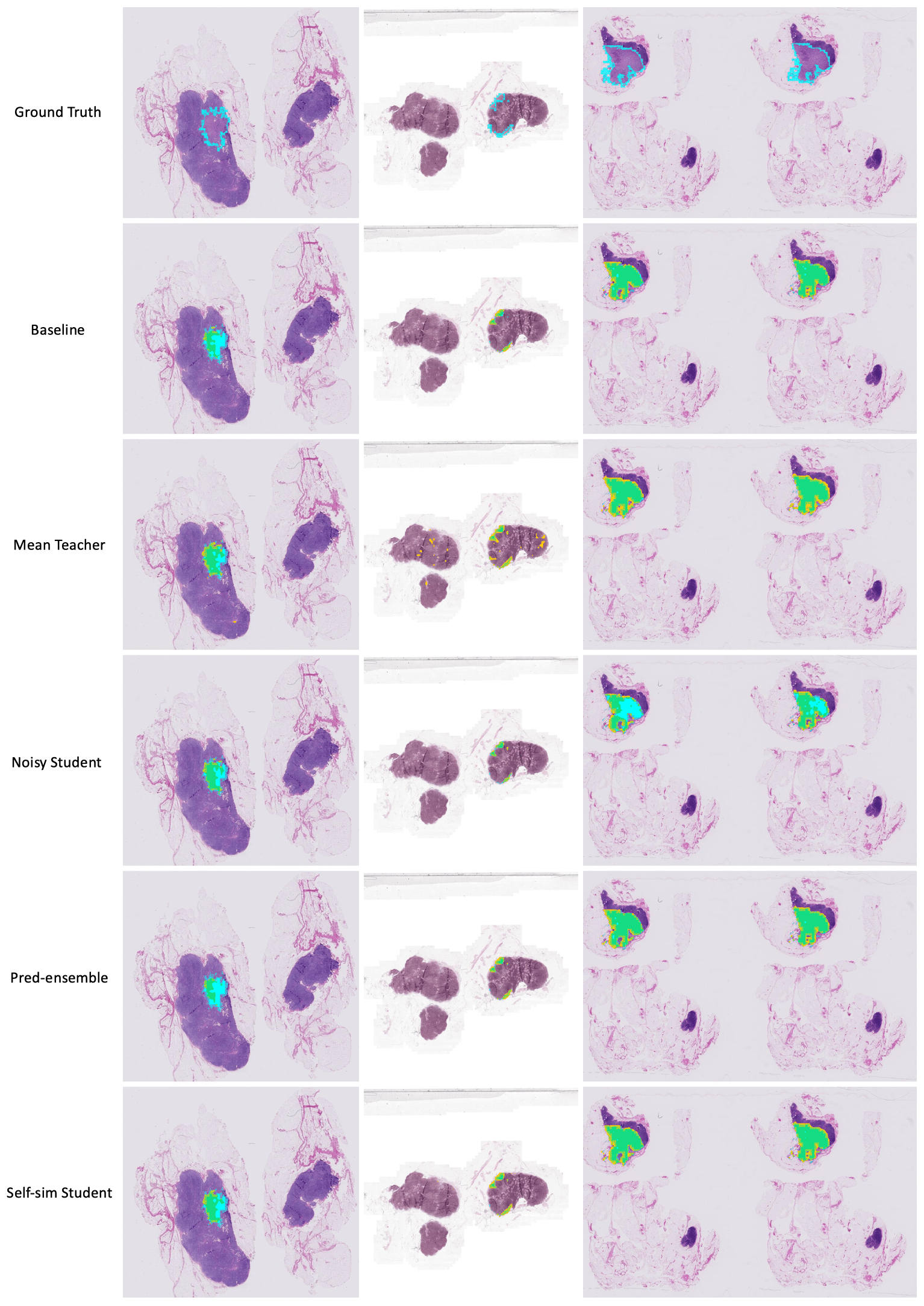}
\end{center}
\caption{
Additional qualitative result on CAMELYON16 dataset. The regions in green color indicate true positives and yellow indicate false positives. The cyan color denotes ground truth (false negatives if no prediction overlapped).}
\label{fig.add_qual_3}
\end{figure}

\begin{figure} 
\begin{center}
\includegraphics[width=1\linewidth]{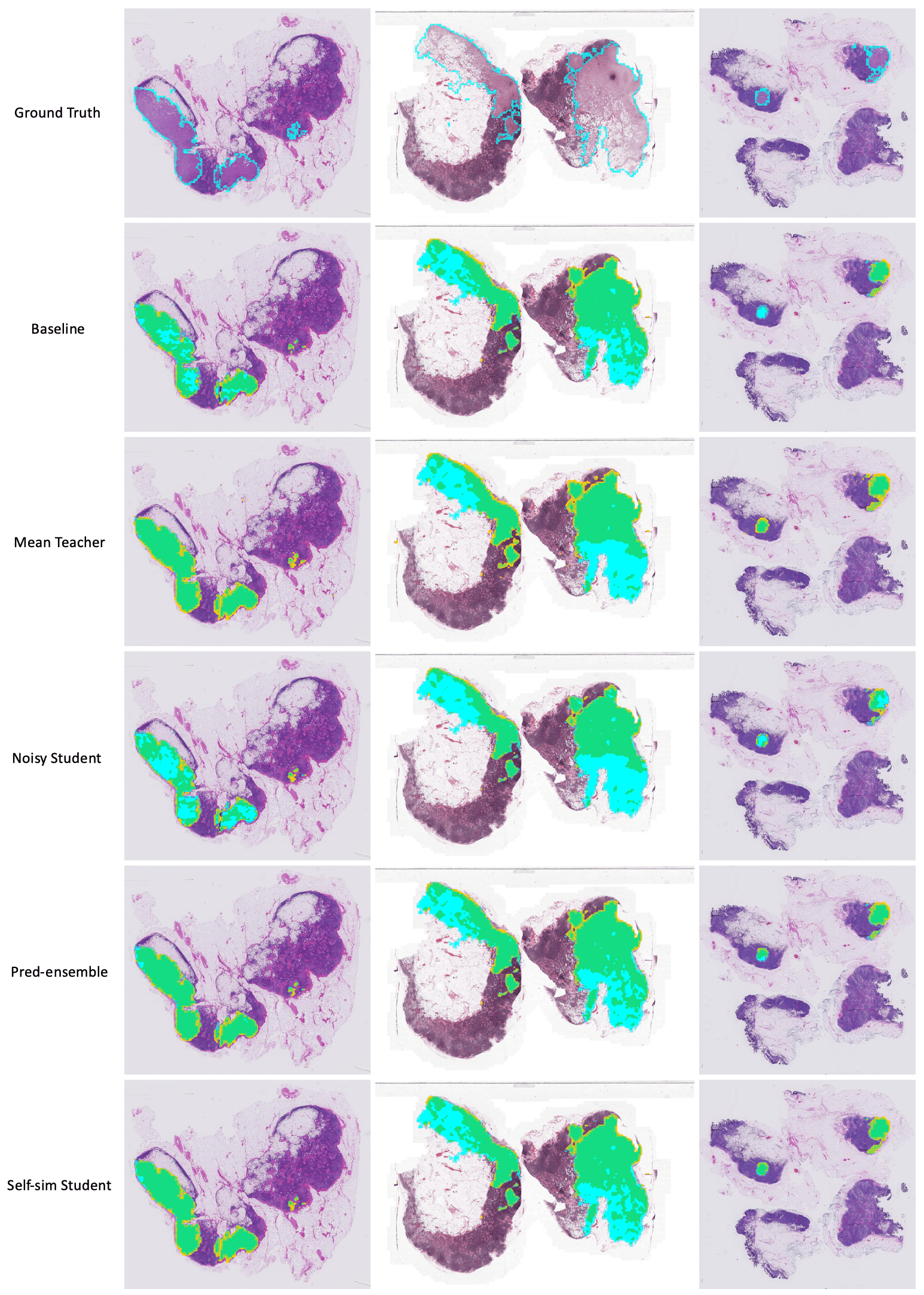}
\end{center}
\caption{
Additional qualitative result on CAMELYON16 dataset. The regions in green color indicate true positives and yellow indicate false positives. The cyan color denotes ground truth (false negatives if no prediction overlapped).}
\label{fig.add_qual_2}
\end{figure}

More qualitative results on TVGH TURP dataset are illustrated in Fig.~\ref{fig.add_turp}. Moreover, the $k_{top} = 1$ results on CAMELYON16 dataset are shown in Fig.~\ref{fig.add_qual_2}, and Fig.~\ref{fig.add_qual_3}. Our Self-sim Student consistently outperforms other previous arts in multiple morphology patterns on TVGH TURP cancer dataset and CAMELYON16 dataset.

\end{document}